\documentclass[11pt]{elsarticle}

\pdfoutput=1

\makeatletter
\def\ps@pprintTitle{%
  \let\@oddhead\@empty
  \let\@evenhead\@empty
  \let\@oddfoot\@empty
  \let\@evenfoot\@oddfoot
}
\makeatother

\usepackage{url}
\usepackage{breakurl}
\usepackage[breaklinks,
            colorlinks = true,
            linkcolor = blue,
            urlcolor  = blue,
            citecolor = blue,
            anchorcolor = blue]{hyperref}

\usepackage{lineno}
\usepackage{textcomp}
\usepackage{graphicx}
\usepackage[margin=1.25in]{geometry}
\usepackage[usenames,dvipsnames]{color}

\usepackage{fancyhdr}
\fancypagestyle{plain}{%
  \fancyhf{}%
  \fancyhead[C]{}
  \fancyfoot[C]{\thepage}
}

\fancypagestyle{empty}{%
  \fancyhf{}%
  \fancyhead[C]{{\it The African School of Physics}}
  \fancyfoot[C]{\thepage}
}
\begin{document}

\begin{frontmatter}


\title{Activity Report on the Ninth African School of Fundamental Physics and Applications (ASP2026)}

\author[add1]{K\'et\'evi A. Assamagan\corref{cor1}}
\ead{ketevi@bnl.gov}
\author[add1]{Mounia Laassiri\corref{cor1}}
\ead{mlaassiri@bnl.gov}
\author[add2]{Bobby Acharya}
\author[add3]{Christine Darve}
\author[add4]{Fernando Ferroni}

\cortext[cor1]{Contacts}

\address[add1]{Brookhaven National Laboratory, USA}
\address[add2]{ICTP, Italy, and King's College London, UK}
\address[add3]{European Spallation Source, Sweden}
\address[add4]{INFN-GSSI, Italy}

\begin{abstract}
\noindent 
The African School of Fundamental Physics and Applications, also known as the African School of Physics (ASP), was initiated in 2010, as a three-week biennial event, to offer additional training in fundamental and applied physics to African students with a minimum of three-year university education. Since its inception, ASP has grown to be much more than a school. ASP has become a series of activities and events to support academic development of African students, teachers, high school pupils (learners) and faculties. We report on the ninth African School of Physics, ASP2026, organized in Kenya, on July 5--19, 2026. ASP2026 included programs for university students, high school teachers and learners.
\end{abstract}


\end{frontmatter}

%



\section{Introduction}
\label{sec:intro}

The African School of Physics is a collection of activities to support academic growths of African students. One activity is a three-week biennial event organized in different African countries---this event consists of a 2-week intensive school, complemented with a one-week African Conference on Fundamental and Applied Physics (ACP)~\cite{ASP2021-reports, ASP, ASP-reports, asp2018, ASP2022, ASP2024}.  The host country of the next biennial event is selected two and half years in advance through a bidding process. In December 2023, Kenya was selected to host ASP2026 at the University of Nairobi and Centre for Mathematics, Science and Technology Education in Africa (CEMASTEA).

In this paper, we present the activity report of ASP2026. In Section~\ref{sec:prog}, we review the scientific program and discuss the supports received in Section~\ref{sec:sup}. We present the profiles of the participants and the expenditures in Sections~\ref{sec:prof} and~\ref{sec:exp} respectively. Feedback from participants are presented in Section~\ref{sec:feed}. Outlook and conclusions are offered in Sections~\ref{sec:out} and~\ref{sec:conc}.

\section{Scientific Program}
\label{sec:prog}

As mentioned in Section~\ref{sec:intro}, The African School of Physics has evolved well beyond three-week biennial engagements. For broader participation in fundamental fields and related applications, the scientific program includes the major physics areas of interest in Africa, as defined by the African Physical Society (AfPS)~\cite{AfPS}:
\begin{itemize}
   \item Particles and related applications: nuclear physics, particle physics, medical physics, (particle)astrophysics \& cosmology, fluid \& plasma physics, complex systems;
   \item Light sources and their applications: light sources, condensed matter \& materials physics, atomic \& molecular physics, optics \& photonics, physics of earth, biophysics;
   \item Cross-cutting fields: accelerator physics, computing, instrumentation \& detectors.
\end{itemize}
Topics in quantum computing \& quantum information and machine learning \& artificial intelligence are also on the agenda. Furthermore, the ASP program includes the fields of societal engagements, namely: topics related to physics education, community engagement, women in physics, early career physicists and engagements with African policymakers in research and education. Representative details on the scientific program are presented in Ref.~\cite{acp2021}, and in the references therein.

The scientific program at ASP2026 was arranged in four distinct components for university students, high school teachers, policymakers, high school pupils (learners). 

The program for university students was set up for the entire duration of the school, July 6--18, 2026. It consisted of lectures, hands-on computing, detectors and instrumentation sessions in the aforementioned topics, organized at the University of Nairobi or Centre for Mathematics, Science and Technology Education in Africa (CEMASTEA). Many activities were arranged in parallel sessions and included online sessions for remote participants~\cite{Students2026}. 

The program for high school teachers was in parallel to the students’ program, during the period of July 6--9 \& 11, 2026. The teachers program was entirely carried out in person at CEMASTEA in plenary or parallel sessions in the topics mentioned above~\cite{Teachers2026}. 

On July 10, 2026, we organized a forum to engage policymakers, where we held—among other sessions—a panel discussion on the 'African Strategy for Fundamental and Applied Physics (ASFAP)' and a dedicated session on 'The need for cultural and structural change within the international physics community to open equitable access and success for Africa.~\cite{Forum2026}.

The learners' program took place on July 13-16, 2026, in parallel with the students' program. During that period, we organized outreach events for over two thousand learners from several high schools in the region of Nairobi.  

Selected photographs of the various activities are compiled in Ref.~\cite{ASP2026-Photos}.

\subsection{Scientific program for students}

Topics of general interests were arranged in plenary sessions during the first week of the school. These were supported by advanced topics developed in parallel sessions. Daily activities started and ended with topics of general interests in plenary sessions. In-between, were combinations of plenary and parallel sessions, to allow for coverage of diverse topics of interest to participants---as presented in Section~\ref{sec:intro}. In the parallel activities, students selected topics that best support the academic majors. In addition, remote participants followed some sessions online.

\subsection{Scientific program for high school teachers}
The objective of the teachers' program is to train teachers in their planning and delivery of physics instructions. The program took place during the week of July 6--9 \& 11, 2026; it consisted of physics plenary lectures complemented by three hands-on activities. The teachers received lectures and hands-on sessions in physics pedagogy, the Internet of Things, physics experimentation, particle physics, accelerator physics  and Astrophysics. Details on the scientific program for high school teachers are in Ref.~\cite{Teachers2026}. A group of 80 Kenyan teachers participated; these were selected nationally by the Ministry of National Education. A joint committee formed by senior officials of the Ministry of National Education and the Local Organizing Committee (LOC) chairs was appointed to review the teachers' program and its objectives, define the selection criteria for teachers, and work on the logistics and financial support needed for a successful program.

\subsection{ASP Forum}
With the ASP Forum, we aim to develop or strengthen strategic engagements with African policymakers in physics education and research, for ASP to better serve the interests of the African community. The detailed agenda of the forum is shown in Ref.~\cite{Forum2026}; it included a panel discussion on the African Strategy for Fundamental and Applied Physics~\cite{asfap}. At the conclusion of the forum, we enjoyed a gala dinner.

\subsection{Scientific program for learners}
The objective for the scientific program for high school pupils is to encourage them to develop and maintain interest in physics. Over two thousand learners from many high schools from the region of Nairobi, Kenya, attended the events in the period of July 13--16, 2026. Four high schools served as the venues, and on different days in July 13--16, 2026, learners from designated high schools converged to a venue. The program was repeated over four days at different venues for different groups of high schools and learners. The participating high schools and learners were selected by the regional education authorities and the school officials. The learners were exposed to physics lectures, demonstrations and hands-on activities in electricity, particle accelerators, and particle detectors.\cite{Learners2026}.

\section{Support}
\label{sec:sup}

We received financial support various institutes\footnote{\url{https://africanschoolofphysics.org/asp2026/}}. Additionally, the University of Nairobi and CEMASTEA provided in-kind logistical and technical assistance. Some lecturers received travel coverage from their institutes.

\section{Participant profiles}
\label{sec:prof}

\begin{figure}[!p]
 \begin{center}
  \includegraphics[width=\textwidth]{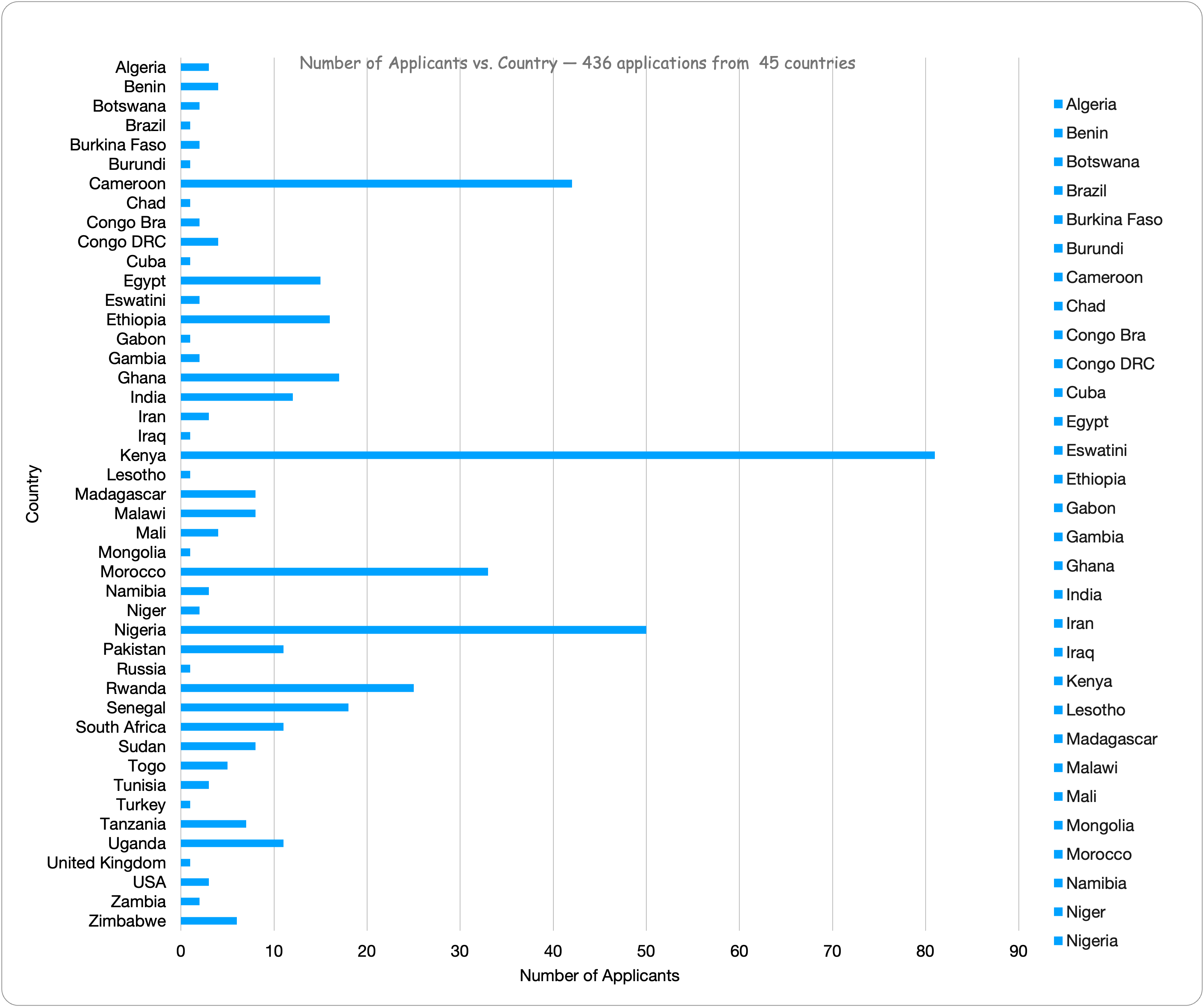}
   \caption{Distribution of ASP2026 student applications as a function of their countries of citizenship.}
    \label{fig:applications}
  \end{center}
\end{figure}

Four hundred and thirty-six (436) candidates from forty-five (45) countries applied for ASP2026 as shown in Figure~\ref{fig:applications}. At the time of writing, it is the largest number of applications received all the edition of ASP. 

\begin{figure}[!htbp]
 \begin{center}
  \includegraphics[width=\textwidth]{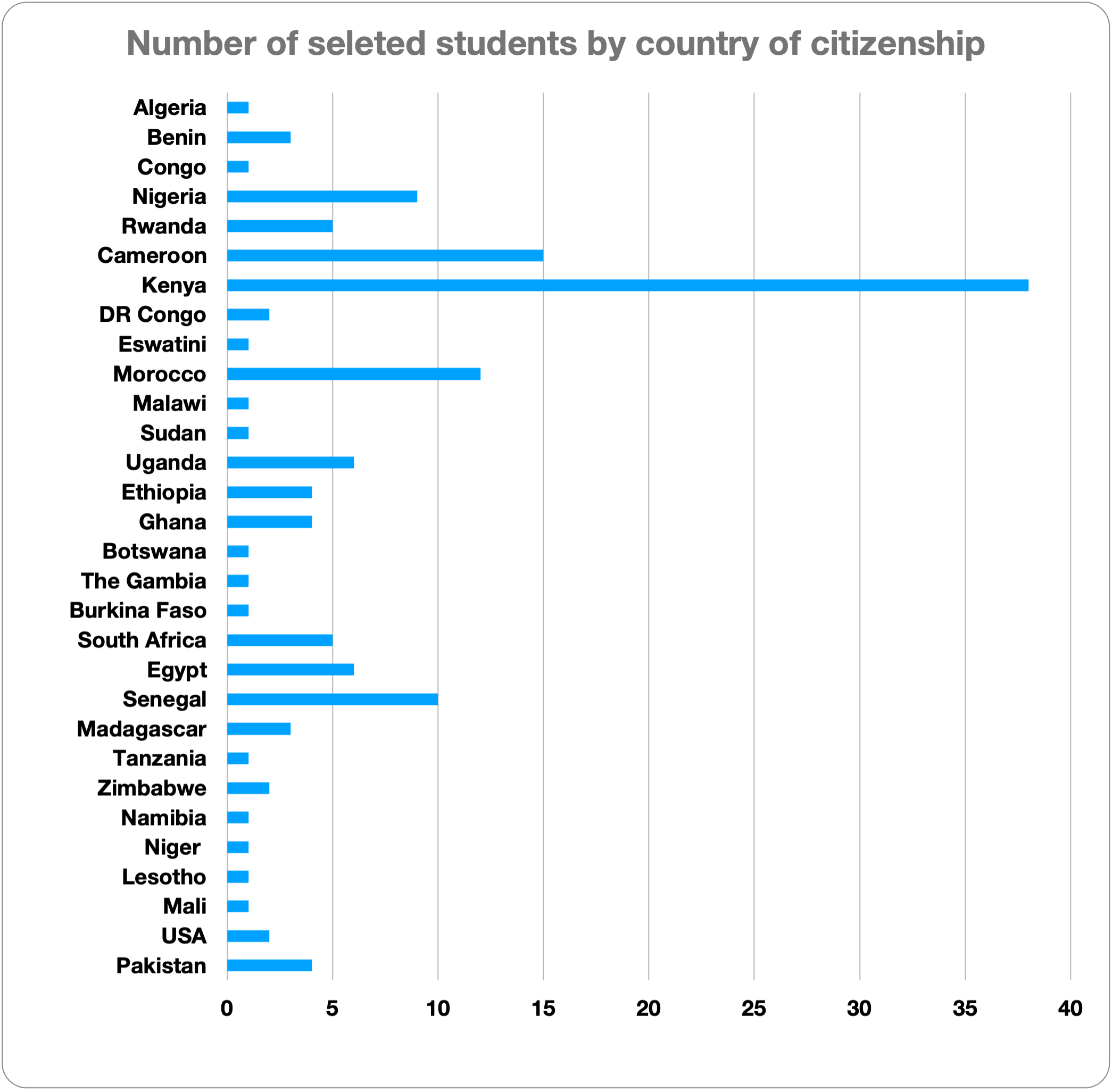}
   \caption{Distribution of selected students for ASP2026 as a function of country of citizenship.}
    \label{fig:selections}
  \end{center}
\end{figure}

All the applications went through a rigorous review process (by an international selection committee of twenty-five members). Applicants had to submit their curriculum vitae, university transcripts, a letter of motivation and arrange for one letter of recommendation. The selection committee (of 16 reviewers) was subdivided into eight subcommittees assigned to review a subset of the applications according to selection criteria set by the International Organizing Committee (IOC), considering requirements from funding agencies. Members of the selection committee were volunteers from the IOC, LOC and lecturers. The distribution of selected students is shown in 
Figure~\ref{fig:selections}.

\begin{figure}[!htb]
 \begin{center}
  \includegraphics[width=\textwidth]{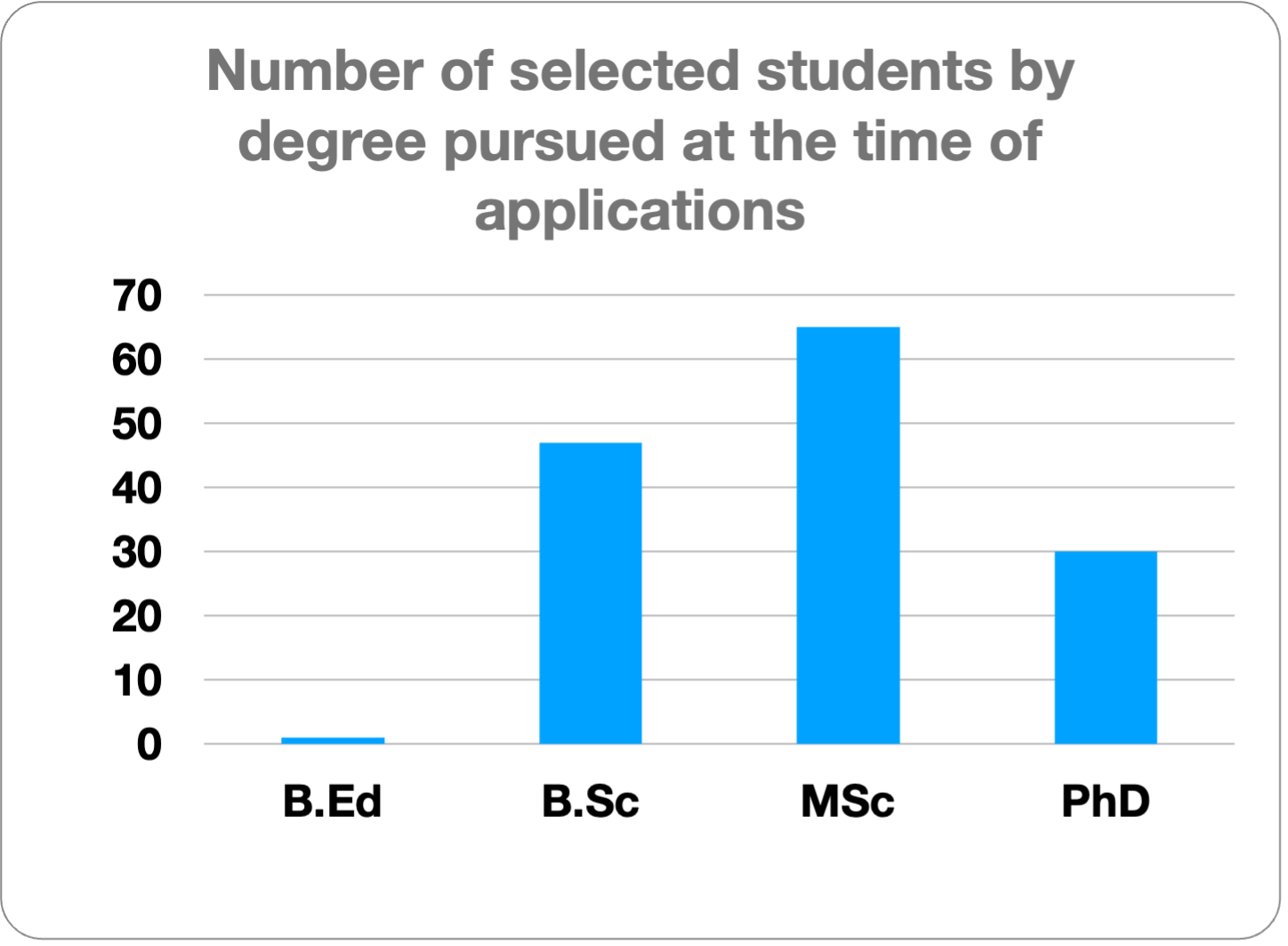}
   \caption{Academic levels of the selected students at ASP2026, at the time of their applications.}
    \label{fig:degrees}
  \end{center}
\end{figure}

\begin{figure}[!htb]
 \begin{center}
  \includegraphics[width=\textwidth]{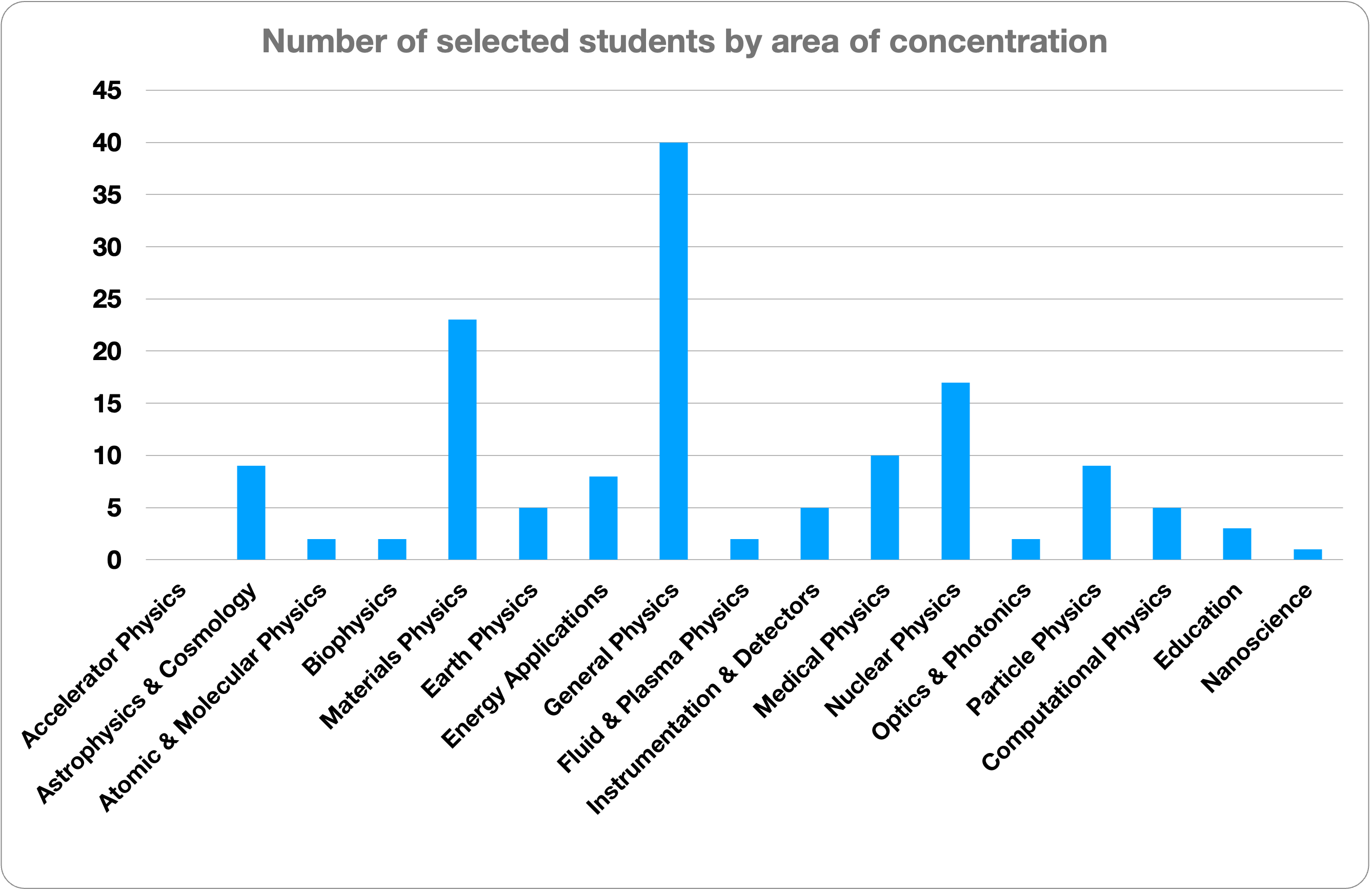}
   \caption{Academic concentrations of the selected students at ASP2026, at the time of their applications.}
    \label{fig:majors}
  \end{center}
\end{figure}

The female-to-male ratio of the selected students was greater than 71:72.  Of the selected students, there were just a few declinations. The selected students were required to have a minimum of three-year university education in engineering, computing, and fundamental and applied physics. Backgrounds on the selected students are shown in Figures~\ref{fig:degrees} and~\ref{fig:majors}.

Eighty high school teachers, selected by the Kenyan education authorities, attended a training workshop, during the period of July 6--9, \& 11, 2026, to improve their skills in the planning and delivery of classroom instructions.

For the high school outreach program, over one thousand  high school learners---of the tenth to the twelfth grades, from many high schools---participated, with an average of about 400 pupils per day during the period of July 13--16, 2026. The time of the high school outreach was chosen to maximize learners' attendance since they were still in school during that period. Four high schools were selected as the venues were selected learners congregated. The activities were repeated daily, for different learners, at a different venue, over the period of four days. the participating learners and high schools were selected by the education authorities of Kenya.

\section{Expenditures}
\label{sec:exp}
The financial support received and described in Section~\ref{sec:sup} was used to cover expenses for ASP2026. These include travels and full room and board for in-person students, local transportation, small detector lab equipment for students, teachers, and pupils.

\section{Feedback}
\label{sec:feed}
Towards the end of the event, we asked participants for feedback. This was carried out in two different approaches. In the first, in-person students were randomly arranged into eleven groups to discuss within their groups and present collective feedback. The second approach was an anonymous survey where participants provided feedback on predefined survey questions.
\subsection{Collective feedback from student groups}
The different groups presented their feedback in a dedicated session on July 17, 2026. From these presentations, we noted the following positive points:
\begin{itemize}
\item ASP2026 offered high quality and passionate lecturers.
\item The organizing team demonstrated great flexibility and adaptability to the needs and suggestions of the participants.
\item There were good interactions between students, lecturers and organizers.
\item ASP2026 offered enhanced skills in the lab and networking opportunities with lecturers and other students.
\item Overall, it was a great and unforgettable experience, during which students learned a lot about physics and its applications, and expanded their networks of friendship and collaboration.
\end{itemize}
There were also many areas where the student groups suggested improvements. We note here the salient ones, to be addressed in future events:
\begin{itemize}
    \item Expand hands-on activities to more areas of physics and applications.
    \item Optimize the range and depth of lectures---plenary and parallel sessions, avoiding repetitive materials---while focusing on the research fields of the participants (and the host country). Preparing a summary of course contents well before the event may help in this optimization.
    \item Do more tests on the program logistics well before the event starts and protection against unsolicited online connections.
    \item Further optimize in the daily programs (coverage and duration), adding practical sessions and group projects to enhance understanding.
    \item Extend the program with field trips, visits to medical or research facilities in the area, and courses on scientific entrepreneurship. 
    \item Improve interactions of online participants.
\end{itemize}

\subsection{Feedback from anonymous survey}
\begin{figure}[!htbp]
 \begin{center}
  \includegraphics[width=\textwidth]{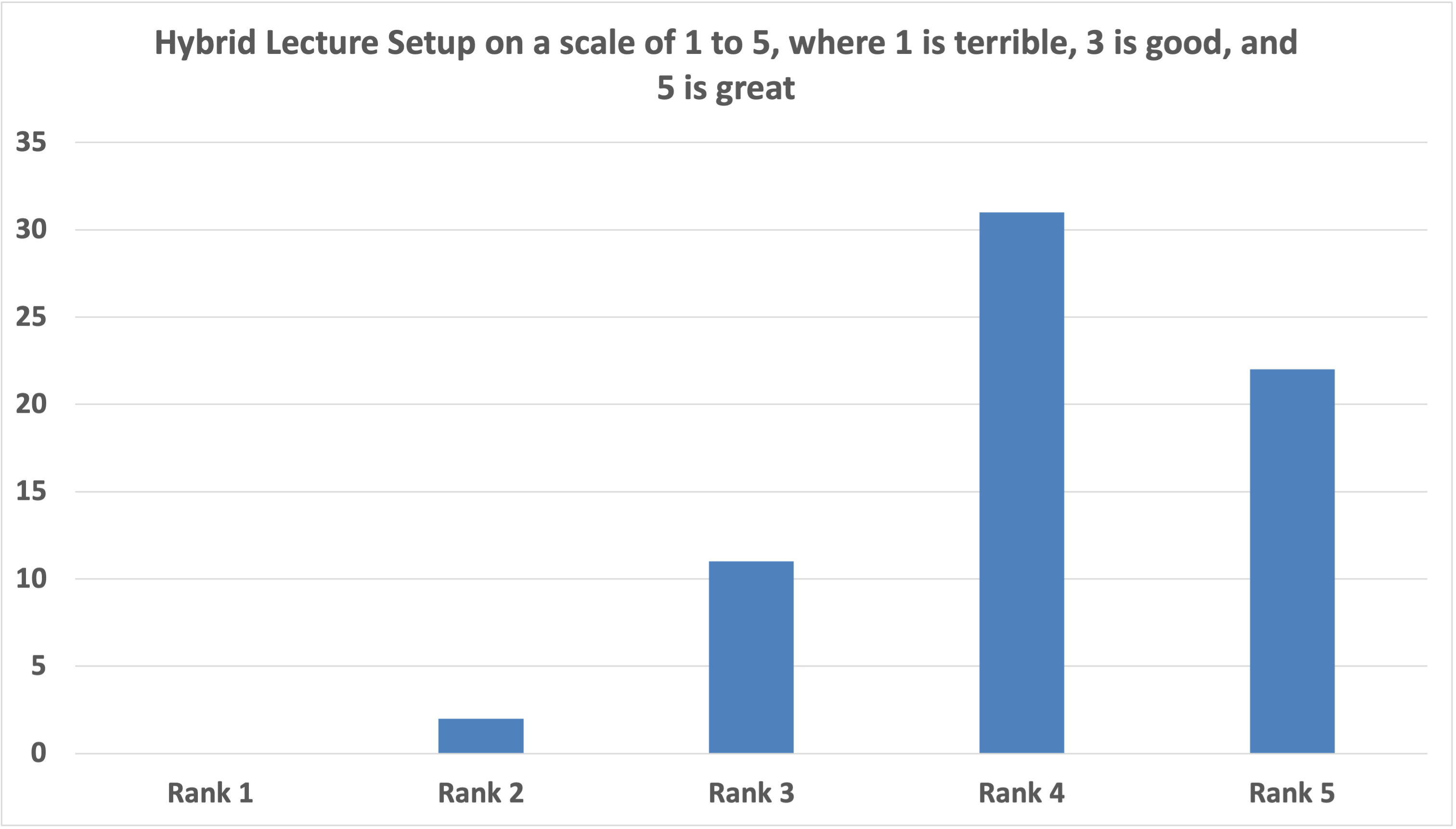}
   \caption{Participants' feedback on the hybrid arrangements.}
    \label{fig:hybrid}
  \end{center}
\end{figure}

\begin{figure}[!htbp]
 \begin{center}
  \includegraphics[width=\textwidth]{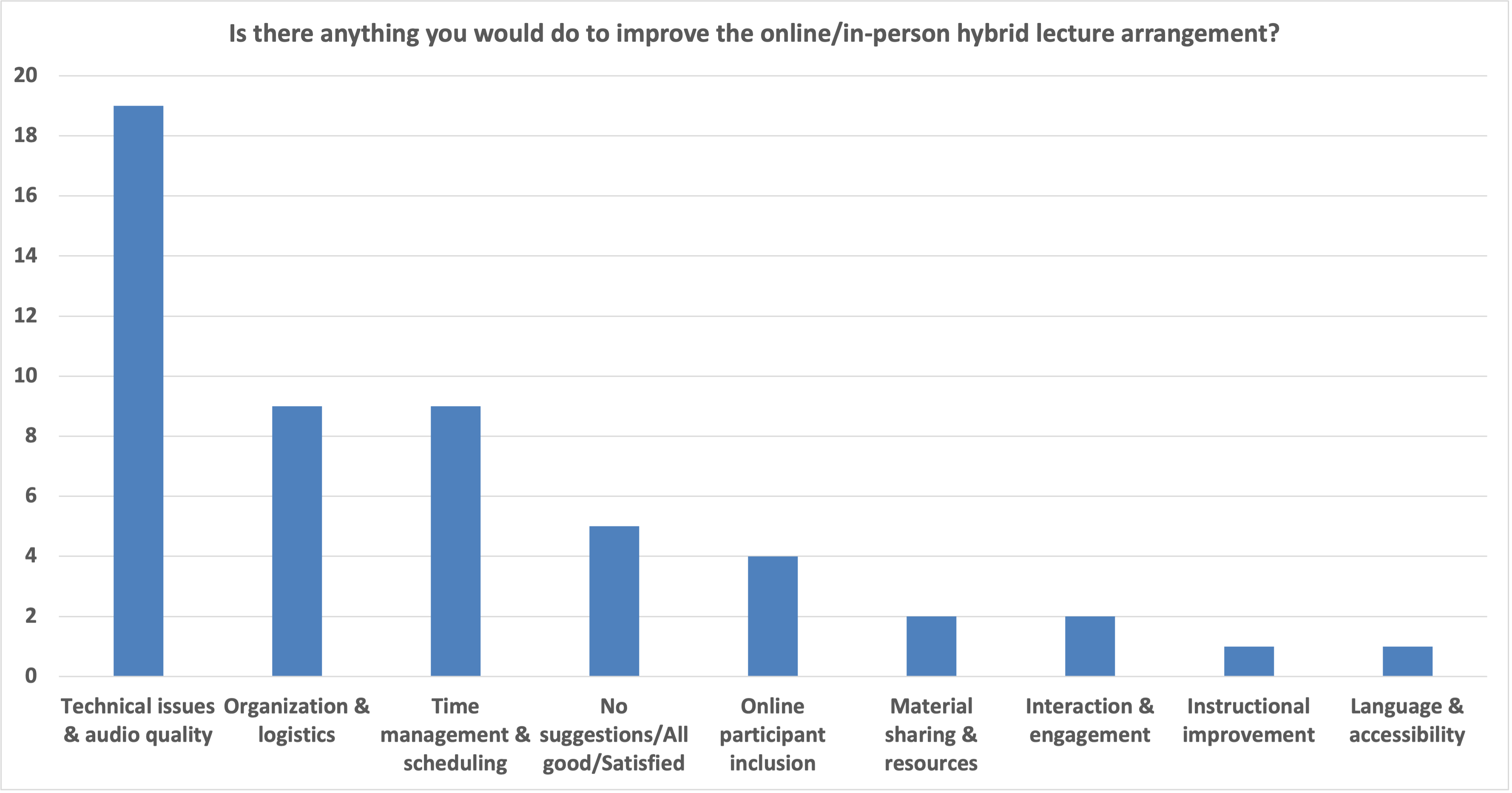}
   \caption{Participants' suggestions to improve the hybrid sessions.}
    \label{fig:hybridImp}
  \end{center}
\end{figure}

Sixty-six participants (in person and online) took the survey. Figures~\ref{fig:hybrid} shows that participants appreciated the hybrid arrangement, although online participants were less satisfied; understandably, most online participants expressed desire for in-person participation. In Figure~\ref{fig:hybridImp}, participants offer suggestions in improving the hybrid sessions.

\begin{figure}[!htbp]
 \begin{center}
  \includegraphics[width=\textwidth]{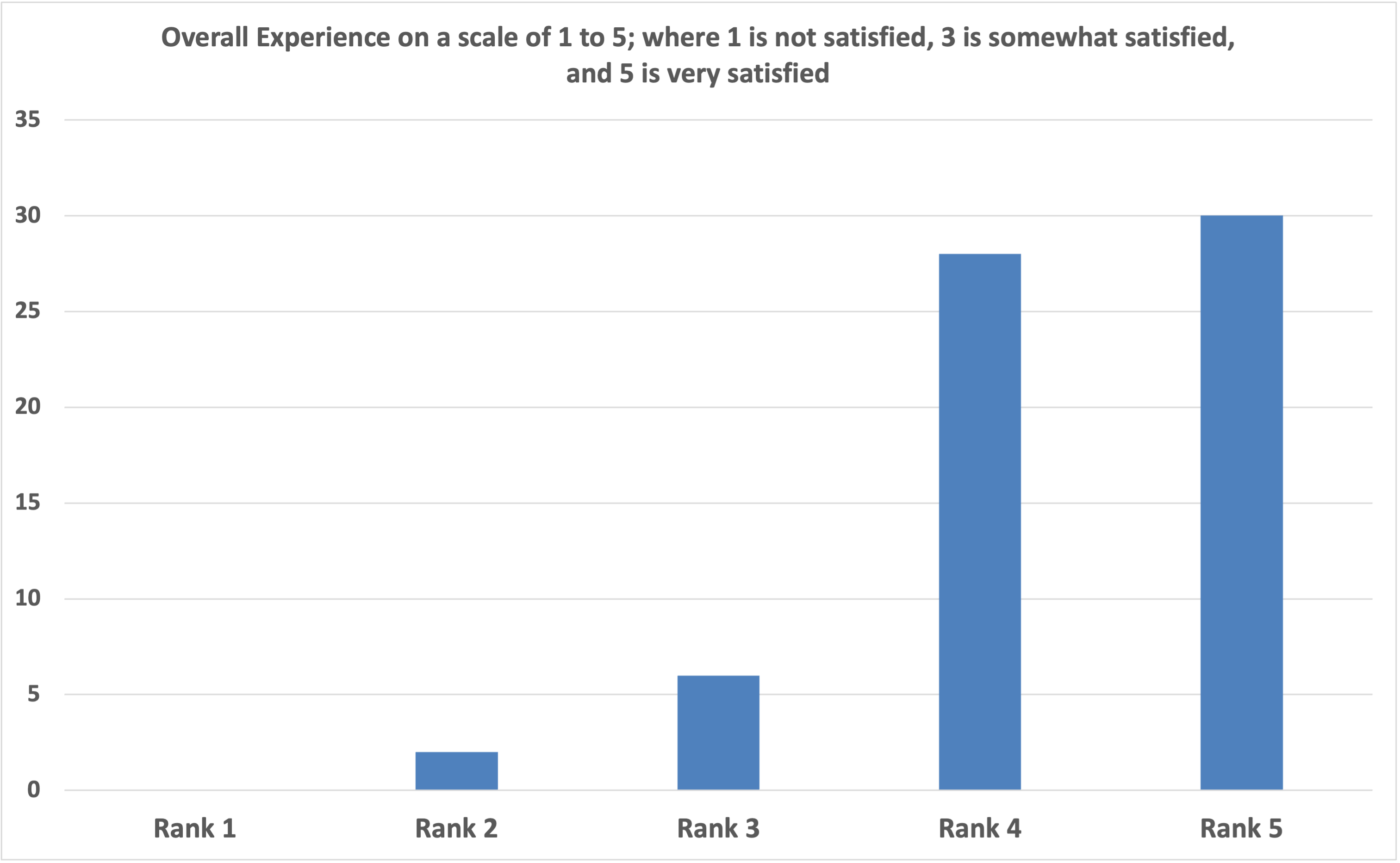}
   \caption{Participants were generally satisfied with their experiences at ASP2026.}
    \label{fig:satisfaction}
  \end{center}
\end{figure}

\begin{figure}[!htbp]
 \begin{center}
  \includegraphics[width=\textwidth]{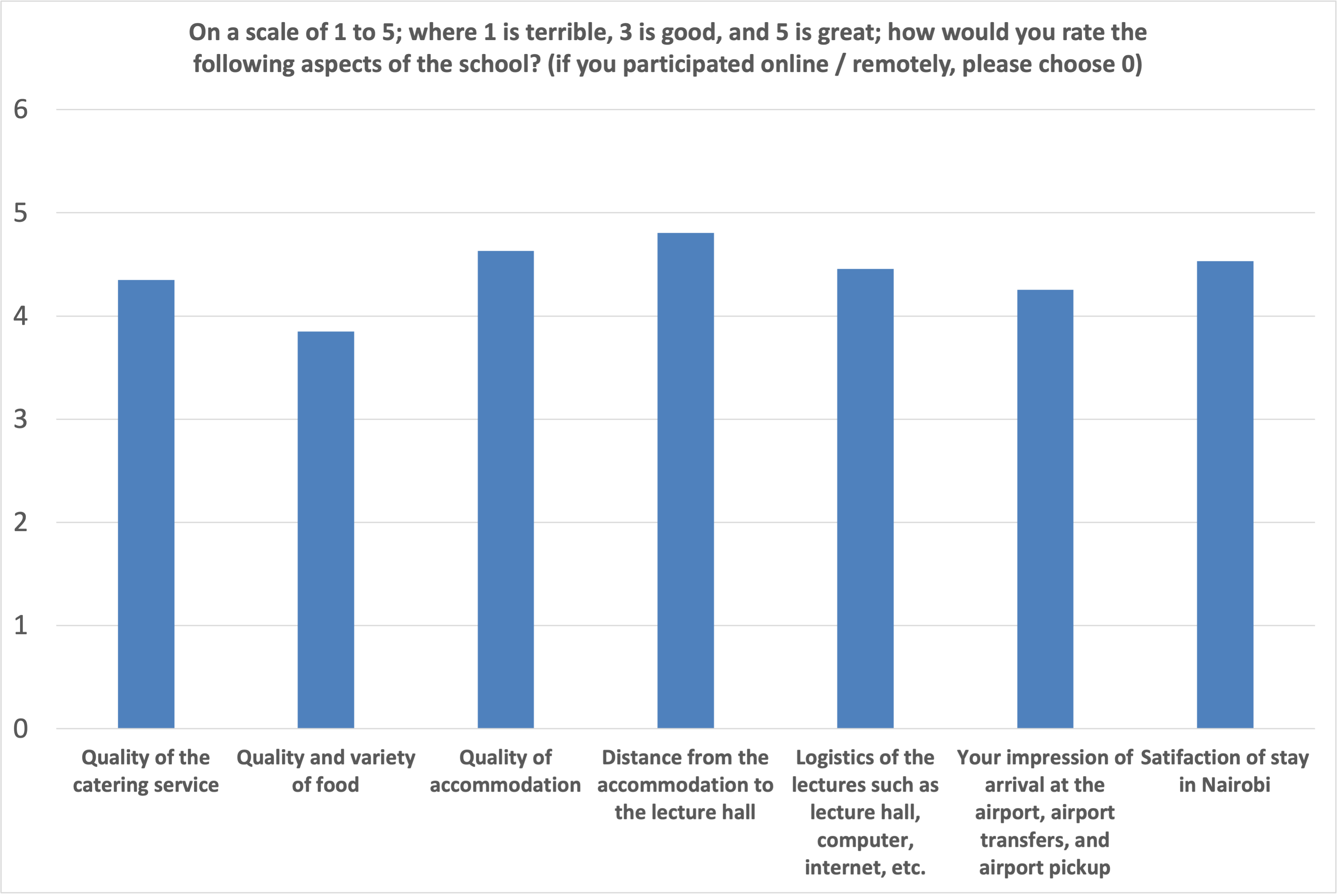}
   \caption{Feedback on various aspects of the logistics.}
    \label{fig:aspects}
  \end{center}
\end{figure}

As shown in Figure~\ref{fig:satisfaction}, most of participants were satisfied with their ASP2026 experiences.
In the details, participants were generally satisfied with various aspects of logistics, as shown in Figure~\ref{fig:aspects}.

\begin{figure}[!htbp]
 \begin{center}
  \includegraphics[width=0.45\textwidth]{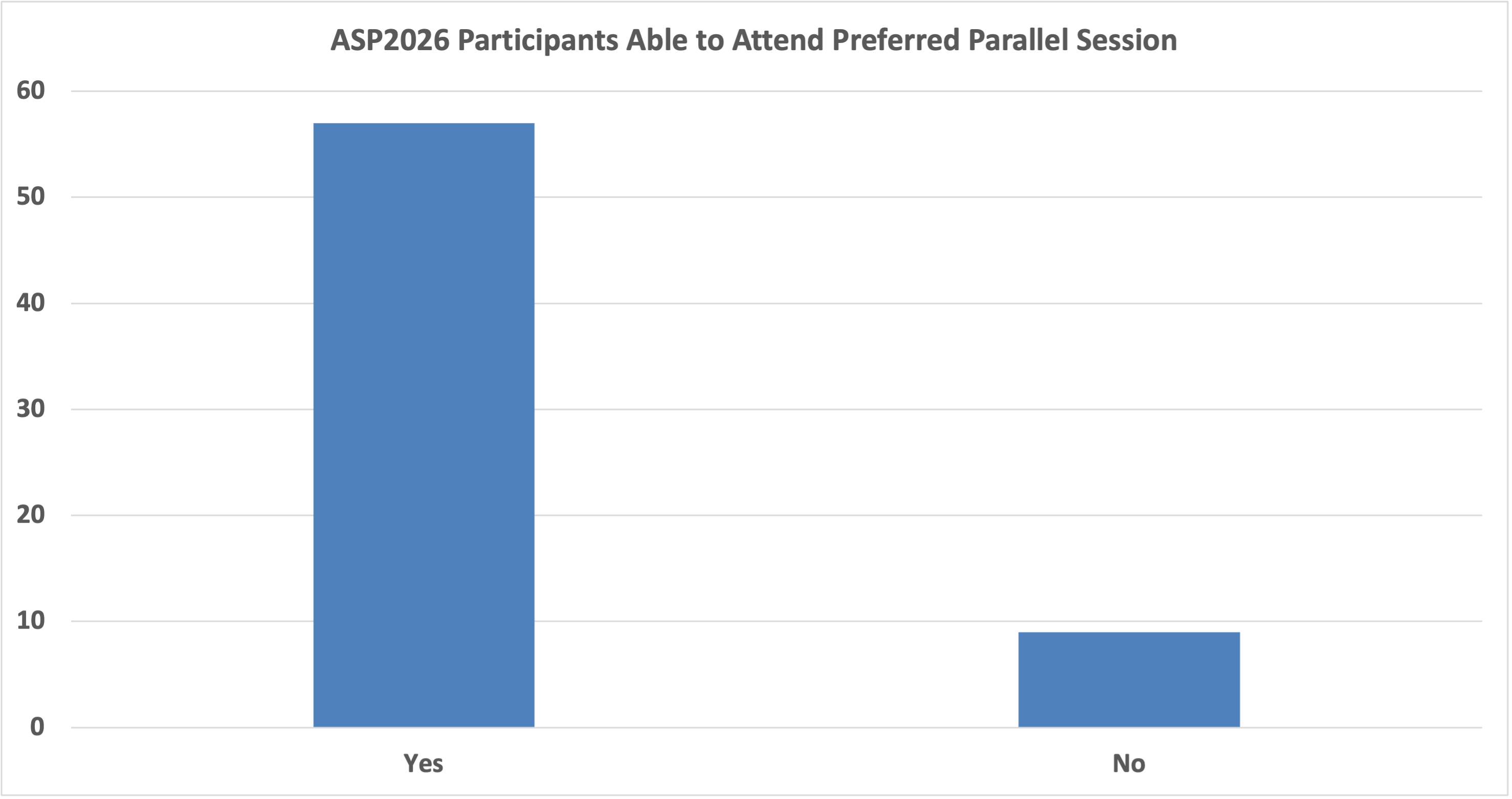}
  \includegraphics[width=0.45\textwidth]{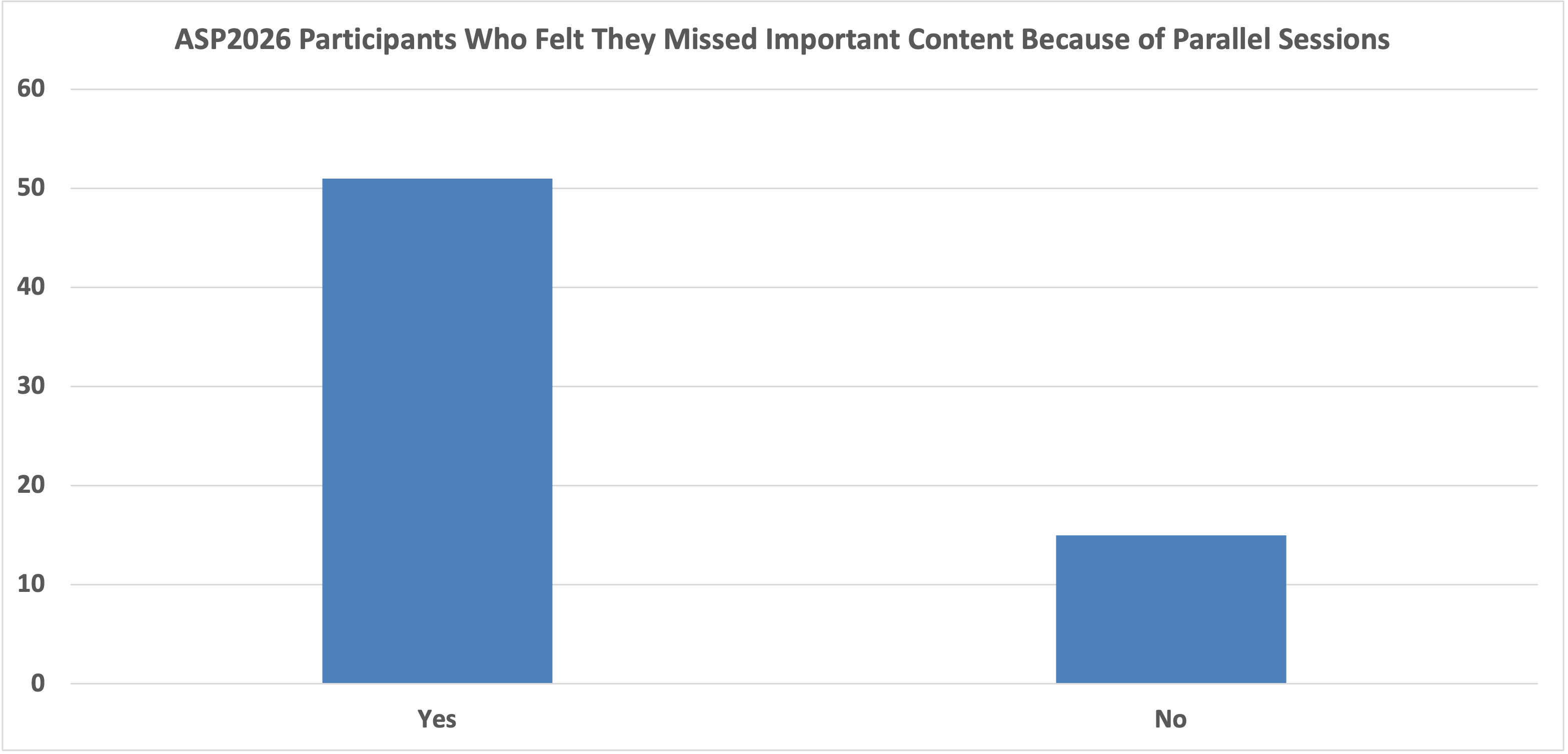}
   \caption{The parallel activities may be improved to cover the interests of all participants more effectively.}
    \label{fig:parallel}
  \end{center}
\end{figure}

\begin{figure}[!htbp]
 \begin{center}
  \includegraphics[width=\textwidth]{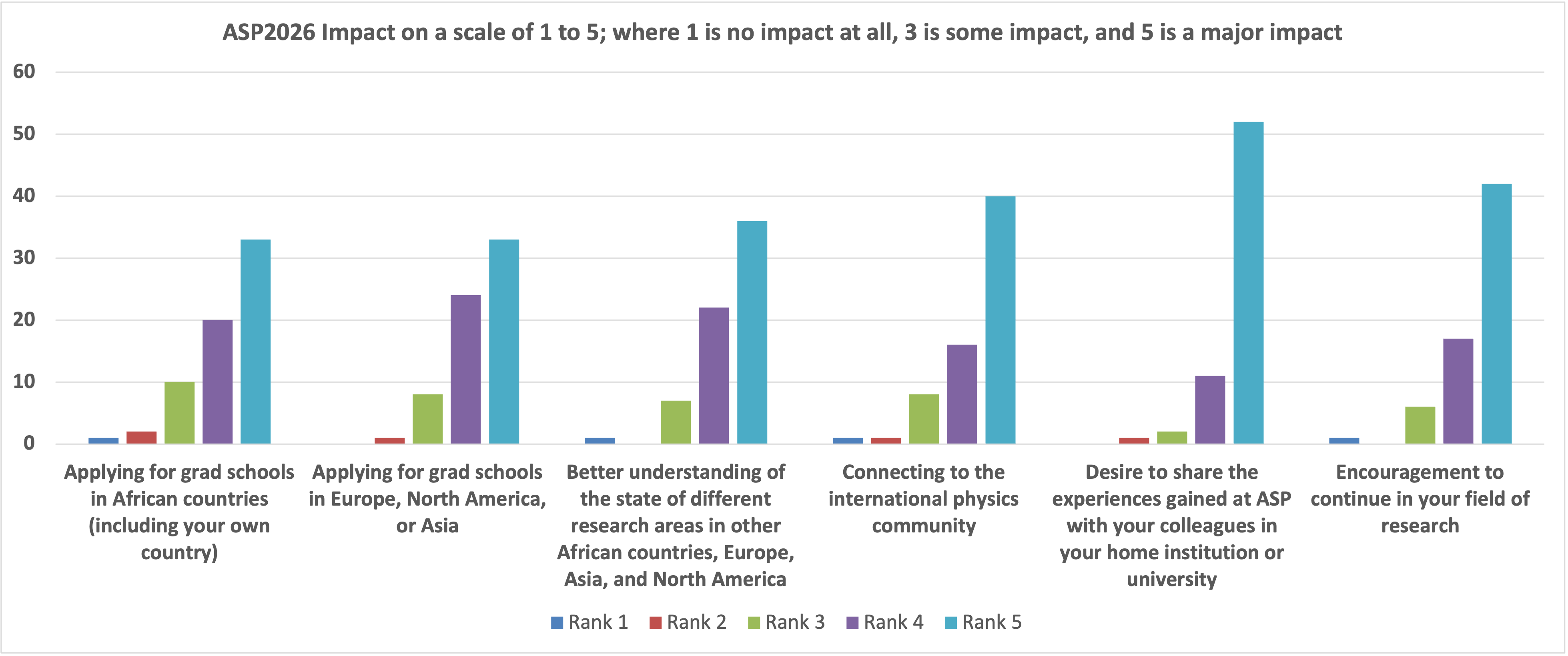}
   \caption{ASP2026 had positive impacts in various aspects of academic engagement.}
    \label{fig:impacts}
  \end{center}
\end{figure}

Most survey respondents said they will recommend ASP to colleagues; however, improvements are needed in the organization of the parallel activities, as shown in Figure~\ref{fig:parallel}.
In Figure~\ref{fig:impacts}, participants reported that ASP2026 had positive impacts on various aspects of their academic efforts.
All the respondents indicated that they would recommend ASP to other students, and expressed interest in fellowship opportunities for higher education. 

General observations for the student survey results are as follows:
\begin{itemize}
\item As expected, the vast majority of survey respondents are from Kenya, going to Kenyan universities.
\item The vast majority of students are studying in their home countries.
\item About half of respondents will graduate in 2026 or next, and the majority are pursuing PhD.
\item About two-thirds respondents were in-person
students and were happy with the hybrid on-line format but there is room for improvement.
\item Similar results for overall satisfaction of ASP
with respect to logistics; the weakest part was the catering and the strongest was the accommodation.
\item The vast majority of students got the desired parallel sessions they wanted, but also felt they missed out on important materials. However, they still support parallel sessions moving forward.
\end{itemize}

Other surveys were conducted for the high school teachers. The teachers also commented favorably on the parallel hands-on activities in their training workshop~\cite{Teachers2026}---most found the sessions at the appropriate level and useful to their own teaching. Most the of teachers who responded to the survey said that they were satisfied with their experience at ASP2026.
\section{Outlook}
\label{sec:out}

Going forward, improvements will be implemented to address feedback from the ASP2026 participants, as presented in Section~\ref{sec:feed}, to the extent feasible.

We look forward to organizing the fifth African Conference on Fundamental and Applied Physics (ACP2026) in Zambia, 2027~\cite{ACP2027}. It will be an international physics conference, with strong participation of ASP alumni and African research faculties to present and discuss their research activities, and form new collaborations. ACP2027 is an activity of ASP, supported through the same funding cycles as the other activities mention in Section~\ref{sec:prog}.

The tenth African School of Physics, ASP2028, is planned in 2028 in Ghana.

\section{Conclusions}
\label{sec:conc}
ASP2026 was organized on July 5--19, 2026, at the University of Nairobi and CEMASTEA in Kenya. It was as a hybrid event. There were over two thousand high school students, eighty high school teachers and 436 student applications, of which about 143 were selected; they were supported by many lecturers and organizers. The scientific program was optimized in a combination of plenary and parallel activities to support the academic growth of the participants. The program was complemented with engagement with policymakers to discuss topics relevant to career development for African students. 

Feedback from participants was generally positive. We noted all the aspects to be improved in future events, namely ACP2027 and ASP2028 as mentioned in Section~\ref{sec:out}. 

\section*{Acknowledgments}
We thank all the sponsors of ASP2026 and the institutes who supported travels for lecturers. We also thank the organizing committee (local and international), in particular Brian Masara (SAIP), Gilbert Tékouté (Executive Management School of Paris, France), the staff at CEMASTEA and the University of Nairobi, Kenya. We acknowledge the efforts of Dr. Sister Mary Taabu, Prof. Kenneth Kaduki and Prof. Zaphania Birech in the local organization of the event. We appreciate the efforts of Sara Sabry (ICTP, SMR4195 -- Secretariat) with the management of student application data. We thank Julia Ann Gray for the student survey, and Kenneth Cecire for the organization of the teachers and learners programs. We are grateful for the tremendous efforts of the lecturers. We acknowledge the collegial atmosphere maintained by all the participants, i.e. students, teachers, lecturers, pupils, organizers and policymakers.

\newpage

\bibliographystyle{elsarticle-num}
\bibliography{ASP2026-Report} 

\end{document}